# Quasi-molecular and atomic phases of dense solid hydrogen

Hanyu Liu, Hui Wang, and Yanming Ma*

*State Key Lab of Superhard Materials, Jilin University, Changchun 130012, China*

The high-pressure phases of solid hydrogen are of fundamental interest and relevant to the interior of giant planets; however, knowledge of these phases is far from complete. Particle swarm optimization (PSO) techniques were applied to a structural search, yielding hitherto unexpected high-pressure phases of solid hydrogen at pressures up to 5 TPa. An exotic quasi-molecular mC24 structure (space group C2/c, stable at 0.47–0.59 TPa) with two types of intramolecular bonds was predicted, providing a deeper understanding of molecular dissociation in solid hydrogen, which has been a mystery for decades. We further predicted the existence of two atomic phases: (i) the oC12 structure (space group *Cmcm*, stable at > 2.1 TPa), consisting of planar $H_3$ clusters, and (ii) the cI16 structure, previously observed in lithium and sodium, stable above 3.5 TPa upon consideration of the zero-point energy. This work clearly revised the known zero-temperature and high-pressure (>0.47 TPa) phase diagram for solid hydrogen and has implications for the constituent structures of giant planets.

**Introduction**

Hydrogen, the first element in the periodic table, is the most abundant material in the interiors of stars and giant planets. High-pressure studies of hydrogen are of fundamental interest and are relevant to our understanding of the physics and chemistry of planetary interiors. At low pressures and temperatures, solid hydrogen (phase I) forms an insulating orientationally disordered phase composed of diatomic molecules. At higher pressures (>0.11 TPa), solid hydrogen is stabilized in an ordered molecular phase II and is transformed into molecular phase III at 0.15 TPa[1-3]. The crystal structure of phase III has attracted much attention[4-8]. Recent theoretical simulations via *ab initio* random structural searches[7] have predicted the best-known candidate structures for phase III. These structures are described by the high-pressure phase sequence C2/c → Cmca-12 → Cmca-8, and above a pressure of 0.49 TPa, molecular hydrogen is predicted to dissociate into atomic hydrogen by adopting a Cs-IV structure. More recently, the random structure searching approach was used by McMahon *et al.* to predict that the Cs-IV structure transits into a triatomic planar *R3m* structure at 2.5 TPa[9]. Of particular interest are the physical mechanisms that underlie the pressure-driven molecular dissociations of solid hydrogen, which remain largely unknown and could shed light on other observed molecular dissociations under pressure (e.g., solid nitrogen, bromine, or iodine) [10-15].

Knowledge of the correct atomic structures of hydrogen could help us assess the earlier suggestion that a room-temperature superconductor may be achievable in atomic hydrogen[16]. The structural studies in Refs.[7,9,17] modified the body-centered cubic structure proposed for atomic hydrogen by Wigner and Huntington in 1935[18]. However, a careful inspection of the enthalpy curves in Ref.[9,17] indicates that the energy landscape of atomic hydrogen is very complex with the existence of many energetically competitive structures. This presents a major difficulty in identifying the best atomic structural models of hydrogen; consequently, additional theoretical efforts are required.

In the present study, we extensively explored the crystal structures of solid hydrogen over the pressure range 0.1–5.0 TPa using first-principles electronic-structure calculations in combination with our recently developed PSO technique for crystal structure prediction[19]. Our goals included (i) an understanding of the physical mechanism underlying molecular dissociation; and (ii) identification of other candidate atomic structures of solid hydrogen to establish the appropriate phase diagrams. The prediction of a quasi-molecular mC24 structure with two different intramolecular bonds allowed us to better understand the molecular dissociations. The results of the structural search in atomic hydrogen require a significant modification of the zero-temperature and high-pressure phase diagrams for solid hydrogen above 2.1 TPa, as we identified two hitherto unexpected atomic structures that are energetically more favorable than the recently proposed atomic $R$3m structures[9].

**Theoretical details**

Our approach is based on a global minimization of free energy surfaces merging *ab initio* total-energy calculations via PSO technique[19] as implemented in CALYPSO code[20]. Our method had several successful application on prediction of high pressure structures of dense Li, Mg, and $Bi_2Te_3$[21-23], among which the blind prediction of insulating Aba2-40 (or oC40) structure of dense Li has been confirmed by independent experiment[24]. The underlying *ab initio* structural relaxations and electronic calculations were performed in the framework of density functional theory within the Perdew-Burke-Ernzerh(PBE)[25] and, as implemented in the VASP (Vienna *ab Initio* Simulation Package) code[26]. The all-electron projector-augmented wave (PAW) method[27,28] was adopted with the PAW potentials taken from the VASP library where $1s^1$ was treated as valence electrons for H atoms, respectively.

**Results and Discussion**

Structural predictions using the CALYPSO (Crystal structure AnaLYsis by Particle Swarm Optimization) code (29) with simulation cell sizes up to 48 atoms/cell were performed over the pressure range 0.1–5.0 TPa. The structural searches successfully recovered all structures predicted earlier in certain pressure ranges. In the low-pressure regime (0.10–0.45 TPa), we predicted $Pca2_1$-8, $P6_3/m$, $P2_1/c$-24, $C2/c$, $Cmca$-12, and $Cmca$-8 structures[7,31], which all possessed typical molecular features. Note that the enthalpy differences of $Pca2_1$-8, $P6_3/m$, and $P2_1/c$-24 were small (<1 meV), indicating that phase II shared similarities with the incommensurate structure found in solid deuterium[32]. Above 0.5 TPa, we identified an exotic quasi-molecular $C2/c$ structure (Pearson symbol mC24) containing 12 molecules per unit cell (Table I and Fig. 1b). Interestingly, this structure possessed a unique feature associated with the intramolecular bonds. Two types of intramolecular bonds were detected. At 0.5 TPa, one type of bond was 0.86 Å in length, whereas the other type was 0.90 Å in length, both much larger than typical H–H bonds (0.74 Å in the gas phase or 0.78 Å in the $Cmca$-8 structure at 0.5 TPa). We therefore regarded the mC24 structure as a quasi-molecular phase. Our enthalpy calculations (Fig. 2a) confirmed the stabilization of the mC24 structure over a wide pressure range, 0.47–0.59 TPa. The identification of mC24 structure revised the earlier dissociation picture from $Cmca$-8 → Cs-IV to $Cmca$-8 → mC24 → Cs-IV. Subsequently, the formation of atomic hydrogen was postponed to 0.59 TPa. The experimental observation of atomic hydrogen, therefore, has been more challenging.

The coexistence of two types of intramolecular bond in the mC24 structure is important for understanding the molecular dissociation of solid hydrogen. Figure 3a shows the pressure-dependence of the intramolecular bond lengths (D1, D2) and a typical intermolecular distance (D3) in the mC24 structure, where D1, D2, and D3 are labeled in Fig. 1b. It was interesting to find that the D2 bond length increased monotonically with pressure, whereas the shorter D1 remained relatively constant. Under pressure, the intermolecular D3 decreased continuously and approached D2 (Fig. 3a). Once D2 and D3 had reached equality (e.g., at 0.6 TPa, D3 was 0.932 Å and D2 was 0.912 Å), the intramolecular and intermolecular bonding became indistinguishable, leading to molecular dissociation. Our current findings resemble the experimental observations of the intermediate I' phase in solid iodine[33] prior to molecular dissociation. This I' phase also displayed two inequivalent intramolecular bonds. By coincidence, the low-pressure phase of mC24 in solid hydrogen and phase I' in solid iodine were both *Cmca*-8 structures (although they differed slightly in their molecular orientations).

Over the pressure range 0.59–2.10 TPa, we successfully predicted the most stable Cs-IV structure, which agreed well with the results of a random structural search [7]. As the pressure increased (>2.1 TPa), we uncovered a group of previously unknown low-enthalpy structures, as shown in Figs. 4(a), 4(b), and 4(d), which were energetically more favorable than the recently proposed *R*3*m* structure [9]. Among these structures, our predicted orthorhombic *Cmcm* structure (Pearson symbol oC12, Fig. 4a) was the most stable one below 5 TPa (Fig. 2a). In contrast with the three-dimensional Cs-IV structure, the oC12 structure assumed a structure with …ABAB… layered stacks consisting of planar triatomic units or $H_3$ clusters. Each $H_3$ cluster may be regarded as an isosceles triangle in which hydrogen atoms occupy the three lattice sites. Two different interatomic distances within each $H_3$ cluster were observed: 0.80 Å and 0.82 Å at 3 TPa (Fig. 4a). These bond lengths were smaller than the nearest H–H distance (0.86 Å at 3 TPa) between two $H_3$ clusters.

Analysis of the electron localization function (Fig. 3b) revealed the remarkable feature that the localized charges were spread out over the entire interstitial areas within the $H_3$ clusters, and the charges weren't localized between different $H_3$ clusters. This result indicated that the $H_3$ clusters could be regarded as $H_3$ molecules, although intramolecular H–H bonding was relatively weak (the electron localization function value fell in the range 0.55–0.64). It should be noted that within each layer of the oC12 structure, the $H_3$ clusters provided nearly perfect packing with alternating opposing orientations, leading to the densest possible atomic arrangement. In contrast, although our predicted *Pnma* (Fig. 4b) and earlier *R*3*m* structures (Fig. 4c) also possessed the structural features of H$_3$ clusters, the orientations were distinct from those in the oC12 structure. As a result, the packing efficiencies in the *Pnma* and *R*3*m* structures were lower. As a further support, the calculated spherical packing efficiency (54.03%) at 3 TPa for the oC12 structure was much larger than the packing efficiencies (in the range 51.41%–51.99%) calculated for the Cs-IV, *Pnma*, and *R*3*m* structures. The Gibbs free energy reduces to H = U + PV at 0 K, where U, P, and V are the static energy, pressure, and volume per proton, respectively. A denser packing structure would reduce the volume, thereby yielding a smaller PV term. Our calculations showed that although the oC12 structure was 31 meV per proton higher in U than the Cs-IV structure at 3 TPa, its PV term was 45 meV lower due to the volume reductions associated with the higher packing efficiency.

Our structural simulations also predicted the existence of the atomic cI16 structure (Fig. 4d), reported here for the first time for solid hydrogen, which is only stable upon inclusion of the zero-point energy (ZPE, see Fig. 2b). This structure was previously observed in the high-pressure phases of lithium[34] and sodium[35,36]. The surprising observation of the cI16 structure in solid hydrogen was not unexpected because hydrogen and lithium (sodium) are both group 1 elements. In some cases, lighter elements can adopt the structures assumed by heavier elements in the same group. The same theory can also apply to the stabilization of Cs-IV in atomic hydrogen. Earlier theoretical studies of paired lithium (Li$_2$) at extreme pressures were conducted in an effort to prepare an H$_2$-like substance[37]. In a different way, the current prediction provides evidence that after molecular dissociation, hydrogen can adopt the high-pressure structures of lithium.

Because hydrogen has a very light atomic mass, its ZPE may be large enough to affect the relative stabilities of its structures. We thus estimated the ZPE for the predicted structures using the quasi-harmonic approximation[38] ($E_{ZPE}=\int d\omega g(\omega)\hbar\omega / 2$). The enthalpies of the various structures, calculated with inclusion of the ZPE as a function of the pressure, are shown in Fig. 2b. At high pressures, inclusion of the ZPE significantly modified the stability field of the oC12 structure and stabilized the cI16 structure above 3.5 TPa. As a consequence, the structural order was modified from Cs-IV → oC12 to Cs-IV → oC12 → cI16.

Analysis of the electronic densities of states (DOS) suggested that the mC24 structure was a very weak metal with a low DOS at the Fermi level (Fig. 5a). This is not surprising because the mC24 structure was a quasi-molecular crystal, and its metallization arose from the pressure-induced band overlap (Fig. 5a). After molecular dissociation, the oC12 and cI16 structures became intrinsic metals and possessed an enhanced metallicity, as evidenced by the large DOS at the Fermi level (Fig. 5b and 5c). This fact agreed with the earlier suggestion that search of room-temperature superconductor in solid hydrogen should be pointed to the atomic phases.

In summary, we applied our recently developed PSO technique for crystal structure prediction toward predicting several candidate structures of solid hydrogen above 0.47 TPa. We unraveled a hitherto unexpected quasi-molecular mC24 structure containing two different intramolecular bond lengths with unusual distinct pressure variations. This result provides an important step toward understanding the pressure-induced molecular dissociation of solid hydrogen, and clearly impacts investigations of similar phenomena in other diatomic–molecular solids, such as oxygen, nitrogen, and the halogens. At higher pressures (>2.1 TPa), we discovered a group of low-enthalpy atomic structures, some of which possess intriguing $H_3$ clusters with various packing orientations. We determined that the oC12 structure, with perfectly opposing orientations of the $H_3$ clusters, provided the best packing efficiency, and the structure was energetically more stable than other such structures. Upon consideration of the ZPE, the cI16 structure identified previously for lithium and sodium became stable relative to the oC12 structure around 3.5 TPa. It has been proposed that the interiors of Saturn and Jupiter are largely composed of hydrogen at pressures up to 4 TPa[39], a pressure regime well within the range studied here. Our findings may improve our understanding of the physics and chemistry in the interiors of giant planets.

**Acknowledgments:**

The authors acknowledge the High Performance Computing Center of Jilin University for supercomputer time and funding from the National Natural Science Foundation of China under grant Nos. 11025418 and 91022029, and the China 973 Program under Grant No. 2011CB808200.

*Author to whom correspondence should be addressed: mym@jlu.edu.cn